\documentstyle[preprint,prb,aps]{revtex}
\tightenlines
\begin{document}



\preprint{Submitted to PR{\bf B}}

\draft

\title{Anomalous Self-Energy Effects of the $\bf B_{1g}$ Phonon  
       in $\bf Y_{1-x}(Pr,Ca)_xBa_2Cu_3O_7$ Films}

\author{A. Bock, S. Ostertun, R. Das Sharma,
             M. R\"ubhausen,\cite{MAR}
             and K.-O. Subke}

\address{Institut f\"ur Angewandte Physik und Zentrum f\"ur 
          Mikrostrukturforschung, \\ 
          Universit\"at Hamburg, Jungiusstra{\ss}e 11, D-20355 Hamburg, 
          Germany}

\author{C.T. Rieck }

\address{I. Institut f\"ur Theoretische Physik,  
          Universit\"at Hamburg, Jungiusstra{\ss}e 9, D-20355 Hamburg, 
          Germany}

\date{\today}

\maketitle

\begin{abstract}

In Raman spectra of cuprate superconductors the 
gap shows up both directly, 
via a redistribution of the electronic background, the 
so-called ``$2\Delta$ peaks'', and 
indirectly, e.g. via the renormalization of phononic excitations.
We use a model that allows us to study the redistribution
and the related phonon self-energy effects 
simultaneously. We apply this model to the $\rm B_{1g}$  
phonon of 
$\rm Y_{1-x}(Pr,Ca)_xBa_2Cu_3O_7$ films, where Pr or 
Ca substitution enables us to investigate under- and overdoped samples.
While various self-energy effects can be explained by the 
strength and energy of the $2\Delta$ peaks, anomalies remain.
We discuss possible origins of these anomalies.

\end{abstract}



\narrowtext

\section{introduction}

Some phonons in high-temperature superconductors show remarkable 
self-energy effects when the superconducting gap $\Delta$ opens.
The strongest effects so far have been observed for modes originating from the 
copper oxygen planes,\cite{Hadjiev1998} in which the pairing mechanism 
is believed to be located.
One example is the $\rm B_{1g}$ 
phonon at 340 cm$^{-1}$ in RE-123 compounds (RE: rare earth or Y) 
where renormalizations have been observed in 
neutron\cite{Reznik1995} 
as well as in Raman experiments.\cite{Friedl1990}
In the latter the $\rm B_{1g}$ phonon has a Fano-type line shape 
as a consequence of the fact, 
that some of the excitations which renormalize it are 
also Raman-active.\cite{Cooper1990}
This holds above $T_{c}$ where quasi-particle scattering leads to a 
structureless incoherent background \cite{Varma1989}
as well as below $T_{c}$ when
the background redistributes and pair-breaking excitations become 
observable.
Considerable theoretical works have been carried out to describe the 
redistribution which leads to the development of so-called 
``$2\Delta$ peaks''.\cite{Devereaux1994,Jiang1996,Strohm1997}
The main features which are observed in the spectra seem to favor 
a $d$-wave gap, however, not all details are understood 
at present.\cite{Manske1998}
For example the fact that the redistribution in $\rm B_{1g}$ symmetry 
weakens with decreasing doping\cite{Nemetschek1997}
indicates that vertex corrections due to strong correlations may play 
a considerable role, as e.g. angle-resolved photoemission spectroscopy 
(ARPES) shows well-defined quasi-particle peaks down to low 
doping levels below $T_c$.\cite{Harris1996}

In this work we investigate the renormalization of the 
$\rm B_{1g}$ phonon and the redistributed background 
in $\rm Y_{1-x}(Pr,Ca)_xBa_2Cu_3O_7$ films
which exhibit $\rm B_{1g}$ $2\Delta$ peaks at 
low temperatures above, near, and below the phonon frequency.
Our aims are first to identify the shape of the electronic background in the 
presence of strongly interacting phonons in order to provide 
substantial data for theoretical descriptions,
and second to correlate the redistribution and the self-energy effects, 
thereby allowing of an investigation of the strength of the electron-phonon 
coupling.
So far, simultaneous descriptions of the $\rm B_{1g}$ phonon and 
the interacting background have been obtained in a
microscopic approach by Devereaux {\em et al.}
\cite{Devereaux1995} and a
phenomenological one by Chen {\em et al.}\cite{Chen1993}
Being mainly interested in the electronic contribution, the latter
authors have taken only a frequency-dependent
imaginary part of the electronic response function into 
account, treating the real part as constant.
As we are interested in the phonon self-energies as well, 
this approximation is not justified here.
The microscopic description considers the full frequency-dependent 
electronic response and provides a basis for the interpretation of 
the fit parameters. 
We will briefly recall that model and show how non-resonant phonon 
Raman scattering, which is to some extent observed in case of the 
$\rm B_{1g}$ phonon,\cite{Heyen1990} 
can be described via a correction of the electron-photon vertex.
We then connect the obtained result 
to the phenomenological Fano formula of 
Chen {\em et al.}\cite{Chen1993} by including a resonant 
excitation channel for the phonon.

The important improvement of our approach is a phenomenological 
description of the frequency-dependent real and imaginary parts of the 
electronic response function which fulfill the Kramers-Kronig 
relations. 
Even though good descriptions of the spectra are obtained,
it turns out that the phonon self-energy effects can only partly be 
assigned to the redistributing background in this model.
Linewidth anomalies, namely a sharpening in the underdoped, an 
intermediate broadening in the slightly overdoped, as well as
strong softenings or hardenings in under- or overdoped samples remain.
We conclude that these additional phonon self-energy effects indicate 
the presence of additional interactions and 
discuss possible explanations for their appearance.

\section{theoretical description}
\label{sec:thde}

As shown by Devereaux {\em et al.}\cite{Devereaux1995} one can 
describe phonon Raman scattering via a correction of the electron-photon 
vertex $\gamma(\mathbf{k})$ plus additional terms arising from 
a photon-phonon vertex $g_{pp}$.
The latter can be regarded as an abbreviation of the resonant 
excitation channel of the phonon involving interband 
excitations.\cite{Hayes1978}
In order to simplify the expression given for the mixed phononic and 
electronic response in Ref. \onlinecite{Devereaux1995}, 
we neglect coupling to interband excitations
in a first approximation.
Figure \ref{figFeynman} shows the Feynman diagram of the considered 
Raman process.
When summing up the bubble diagrams
one obtains the following expression for the full response function
$\chi_{\sigma}(\omega)$ in 
symmetry channel $\sigma$ of the lattice:
\begin{equation}
  \chi_{\sigma}(\omega)= \gamma_{\sigma}^2\chi^e_{\sigma}(\omega) -
  \gamma_{\sigma}^2g_{\sigma}^2\chi^e_{\sigma}(\omega)^2 
  D_{\sigma}(\omega) \,,
  \label{eqResponse}
\end{equation}
where $\gamma_{\sigma}$ represents the symmetry elements of the 
Raman vertex projected out by the incoming and outgoing polarization 
vectors and 
$g_{\sigma}$ is the lowest order expansion coefficient of the 
electron-phonon vertex which is independent of momentum but 
may vary with temperature.\cite{Devereaux1995}
$\chi^e_{\sigma}(\omega)=
R^e_{\sigma}(\omega) +i\varrho^e_{\sigma}(\omega)$ 
is the electronic response, whose real and imaginary parts are connected by 
Kramers-Kronig relations,
$D_{\sigma}(\omega)=D^0_{\sigma}(\omega)/
[1+g_{\sigma}^2\chi^e_{\sigma}(\omega)D^0_{\sigma}(\omega)]$ is the 
renormalized, and
$D^0(\omega)=2\omega_p/(\omega^2-\omega_p^2+2i\omega_p\Gamma)$  
the bare phonon propagator. 
The latter contains the bare phonon frequency $\omega_p$ and the bare 
phonon linewidth $\Gamma$ (HWHM). 
To simplify Eq. (\ref{eqResponse}) further,
we make the assumption that the phonon is only 
renormalized by the Raman-active electronic response 
$\gamma_{\sigma}^2\chi^e_{\sigma}(\omega)$, 
neglecting all other self-energy contributions, 
e.g. those due to anharmonic phonon-phonon interactions.
They are therefore included in $\omega_p$ and $\Gamma$.

The fluctuation-dissipation theorem relates the Raman intensity 
$I(\omega)$ to $\left[1+n(\omega,T)\right]
{\rm Im}\,\chi_{\sigma}(\omega)$
with $n(\omega,T)=1/\left[\exp(\hbar\omega/k_BT)-1\right]$ the Bose 
factor.
If $A$ is the proportionality constant between the Raman efficiency
$I_{0}(\omega)=I(\omega)/[1+n(\omega,T)]$ and 
${\rm Im}\,\chi_{\sigma}(\omega)$ 
then the constant $C=A\gamma_{\sigma}^2/g_{\sigma}^2$ allows us to express the Raman 
efficiency in the following way:
\begin{displaymath}
	I_0(\omega)=\varrho_*(\omega)+\frac{C}{\gamma(\omega)
	\left[1+\epsilon^2(\omega)\right]}  \: \times 
\end{displaymath}
\begin{eqnarray}
  \left\{ \left[\frac{R_*(\omega)}C\right]^2 - 
  2\epsilon(\omega)\frac{R_*(\omega)}C\frac{\varrho_*(\omega)}C - 
  \left[\frac{\varrho_*(\omega)}C\right]^2 \right\} \, , 
\label{eqRamInt}
\end{eqnarray}
with the substitutions 
$\varrho_*(\omega)=Cg_{\sigma}^2\varrho^e_{\sigma}(\omega)$, 
$R_*(\omega)=Cg_{\sigma}^2R^e_{\sigma}(\omega)$,
$\epsilon(\omega)=\left[\omega^2-\omega^2_{\nu}(\omega)\right]
/2\omega_p\gamma(\omega)$,
$\gamma(\omega)=\Gamma+\varrho_*(\omega)/C$, and
$\omega_{\nu}^2(\omega)=\omega_p^2-2\omega_pR_*(\omega)/C$.

It turns out that Eq. (\ref{eqRamInt}) is similar to the 
phenomenological Fano formula that Chen {\em et al.}\cite{Chen1993} 
obtained using an extended Green's function model,
except that the phononic signal vanishes here when 
the electron-phonon coupling constant is zero.
This is related to the fact that we have neglected the resonant 
excitation channel of the phonon so far.
Therefore, one would expect a decreasing phonon intensity if the 
background decreases, or if the system becomes insulating.
Examples for both cases are, e.g. the strong Fano-type profiles of the 
$\rm A_{1g}$ plane-oxygen phonons in 
$\rm HgBa_2Ca_3Cu_4O_{10+\delta}$,\cite{Hadjiev1998} which almost 
vanish above $T_c$ along with the background feature, 
or the Fano-type Ba mode in $xx$ or $yy$ polarization in 
$\rm YBa_2Cu_3O_{6+y}$ (Ref. \onlinecite{Burns1991}) which is absent in the 
antiferromagnetic compound at y=0.
In case of the here investigated $\rm B_{1g}$ phonon, however, it is 
known that substantial intensity is observed even at low oxygen 
contents.\cite{Burns1991} 
In order to connect Eq. (\ref{eqRamInt}) to the  
phenomenological Fano formula of Ref. \onlinecite{Chen1993} 
the replacement 
$R_*(\omega) \rightarrow R_*(\omega)+R_0$
in the numerator of the second term is sufficient.
Hereafter we will refer to the modified equation as 
Eq. (\ref{eqRamInt},$R_0$). 
The additional fit parameter $R_0=Cg_{\sigma}^2R_{pp}$
describes the intensity of the phonon originating from the resonant 
excitation channel.\cite{R_0}
With this parameter one keeps a Lorentzian phonon 
at $\omega_p$ for vanishing 
effective electron-phonon coupling $g_{\sigma}$ 
if $g_{\sigma} \cdot R_0$ remains finite. 
This replaces the $q^2 \cdot C_{Fano} $-condition of simple Fano 
profiles where 
$I_{Fano}(\omega)=C_{Fano} \cdot (q+\epsilon)^2/(1+\epsilon^2)$ with 
$\epsilon =(\omega- \omega_{\nu})/\gamma$ and $q$ the asymmetry 
parameter.\cite{Cooper1990}
Physically that means that one obtains a Lorentzian profile with 
Eq. (\ref{eqRamInt},$R_0$) when the phonon solely couples to electronic 
excitations which are strongly peaked away from the phonon energy.
In reference to the simple Fano approach, in which the total 
phonon intensity is given by 
$I=\pi \gamma C_{Fano} q^2$, \cite{intensity}
we find for the total and the bare phonon intensity 
$I_{tot}=\frac{\pi}{C} [R_*(\omega_p)+R_0]^2$ and 
$I_{phon}=\frac{\pi}{C} R_0^2$.

Our aim is to obtain the self-energy effects in a consistent way by 
using a phenomenological description of the electronic background.
For this, we express the measured electronic response 
$\varrho_*(\omega)$ with three terms: 
\begin{displaymath}
  \varrho_*(\omega)=
  I_{\infty}\tanh\left(\omega/\omega_T\right) + 
\end{displaymath}
\begin{equation}
  \left[\frac{C_{2\Delta}}{1+\epsilon_{2\Delta}^2(\omega)} - 
  \left(\omega \rightarrow-\omega \right) \right] -
  \left[\frac{C_{sup}}{1+\epsilon_{sup}^2(\omega)} - 
  \left(\omega \rightarrow-\omega \right) \right]  \: .
  \label{eqRho}
\end{equation}
The first one describes the incoherent background,\cite{Varma1989} 
the second contribution models the $2\Delta$ peak, 
and the third one the suppression of spectral weight seen at low Raman 
shifts. 
$I_{\infty}$, $C_{2\Delta}$, and $C_{sup}$ are fit parameters for the 
respective intensities.
Two-magnon excitations are not explicitly included in this 
description.
The crossover frequency $\omega_T$ of the background,
which is cut off at $\omega_{cut}=8000$ cm$^{-1}$,
is used as a fit parameter, however, allowing only values close to the 
spot temperature.
For the Lorentzians the abbreviations
$\epsilon_j(\omega)=(\omega-\omega_j)/\Gamma_j$
with $j=2\Delta,sup$ are used. 
The second terms in the brackets are necessary to fulfill the 
symmetry requirements for the Raman response. 
The Lorentzian modelling the suppression 
is arbitrarily bound to the other one by:
$2\omega_{sup}=2\Gamma_{sup}=\omega_{2\Delta}-\Gamma_{2\Delta}$ in 
order to reduce the number of free parameters.
Eq. (\ref{eqRho}) is an appropriate way to model the imaginary 
part of the electronic response,
however, the real part of the response function
contains a constant error as a result of the arbitrarily chosen cut-off 
frequency.\cite{KKRvonTanH}
This error will modify the self-energy $R_*(\omega_p)/C$ 
and, as a consequence, the resonant intensity contribution $R_0$.
Both will hence depend on the chosen value of $\omega_{cut}$.
Analysis of the Ba mode in $\rm A_{1g}$ polarization confirms us that 
this error is small as this mode can almost completely be described by the 
interaction with the background independent of 
$\omega_{cut}$.\cite{Bock1999}
In detail, it turns out that 
$R_*(\omega_p)/C$ \{$R_0/C$\} increases \{decreases\} by
$I_{\infty}/C \cdot \frac{2}{\pi}\mathrm{ln}$
$(\omega_{cut}^{new}/\omega_{cut}^{old})$ when the 
cutoff frequency is varied.
With a typical value of $I_{\infty}/C=1$ in the present study,
corrections of $\sim 0.2$ cm$^{-1}$ appear,
when $\omega_{cut}$ is 
increased from 8000 cm$^{-1}$ to 11000 cm$^{-1}$.
Whereas the absolute values of the bare phonon frequency 
and the bare phonon intensity
have to be considered with some care for the reason given above,
the relative changes at different temperatures are not affected.

Even though all three terms in Eq. (\ref{eqRho}) are needed to 
describe the background below $T_c$, the last two terms  
diminish continuously close to the transition into the normal state.
Above the transition, the background is almost entirely described by 
the incoherent contribution except for a small broad hump at the 
position where the $2\Delta$ peak was observed close to $T_c$. 

\section{experimental details}
\label{sec:exp}

We study $\rm Y_{1-x}(Pr,Ca)_xBa_2Cu_3O_7$ films 
with different doping levels achieved by substituting Y 
partially with Pr (underdoping) or Ca (overdoping).
The investigated films are a 
x=0.1 Pr-doped film 
($\rm Y_{0.9}Pr_{0.1}$), a pure x=0 film (Y-123), and a 
x=0.05 Ca-doped film ($\rm Y_{0.95}Ca_{0.05}$) abbreviated in the 
following as given in brackets.
They have resistively measured $T_c$'s,
as defined by zero resistance,
of 86.3\,K, 88.0\,K, and 82.7\,K, respectively.
The films are grown by pulsed laser deposition 
on SrTiO$_3$ substrates in a process optimized to obtain high oxygen 
contents.\cite{Dieckmann1996}
Whereas the growth of the $\rm Y_{1-x}Pr_xBa_2Cu_3O_7$ films has also
been optimized to obtain homogeneous and smooth films with low precipitate 
densities, the Ca-doped film
exhibits a somewhat poorer surface quality, and a local variation of 
the Ca-content from the nominal value cannot be excluded.
However, also for this film
the temperature where the superconductivity-induced features 
vanish is in good agreement with the measured $T_c$.
All films are $c$-axis oriented and their
high degree of in-plane orientation allows us to study the 
plane-polarized excitation geometries.\cite{Dieckmann1996}
Raman spectra have been taken using the 458 nm Ar$^+$ line
in a setup described elsewhere.\cite{Bock1998}
They are corrected for the response of spectrometer and detector.
We have used a power density of 110 Wcm$^{-2}$ and an
effective spot radius of 40 $\mu$m leading to a typical heating  
of 2 K in the laser spot.\cite{Bock1995} 
All given temperatures are spot temperatures.
In order to measure spectra with $\rm B_{1g}$ symmetry,
the incident light was polarized along $x'=(1,1,0)$ and 
the scattered light along $y'=(1,-1,0)$ in a coordinate system given 
by the crystal axes of the substrate.

\section{results}
\label{sec:result}

In Fig.\,\ref{figDopingRaman} we show Raman spectra of the
films below and above $T_c$.
For a better comparison of the spectra we have normalized the 
intensity of the film Y-123 to unity above 650 cm$^{-1}$ and used the 
same scaling factor for the other films as well.
This factor is also used in the following figures.
The spectra are efficiency corrected for the thermal factor
$\left[1+n(\omega,T)\right]$.
The electronic continua, which are shown in the insets, are obtained 
after subtraction of the phonons. 
Whereas this can easily be done in case of the weak Ba and symmetric 
O(4) modes, the description given in Eq.~(\ref{eqRamInt},$R_0$) is 
used to subtract the $\rm B_{1g}$ mode properly.
This description is also used for the O(4) mode in 
$\rm Y_{0.9}Pr_{0.1}$ and for the Ba mode in $\rm Y_{0.95}Ca_{0.05}$.
In all films we observe a featureless background at 152\,K which 
becomes almost constant above 400 cm$^{-1}$. 
At 18\,K the $\rm B_{1g}$ $2\Delta$ peaks are visible as well 
as suppressions of spectral weight at small Raman shifts.
Whereas the $\rm B_{1g}$ $2\Delta$ peak can clearly be seen in the raw data of 
$\rm Y_{0.95}Ca_{0.05}$ they are somehow hidden in the spectra of the 
films with lower doping levels.
In $\rm Y_{0.9}Pr_{0.1}$ this results from the strongly 
reduced intensity of the redistribution
(note the scale in the insets of Fig. \ref{figDopingRaman})
whereas in Y-123 it is a 
consequence of the destructive interference of phonon and background 
right at the position of the $\rm B_{1g}$ $2\Delta$ peak.
In agreement with recent studies\cite{Nemetschek1997,Chen1993}
we find that the redistribution becomes weaker at low 
doping levels and that the $\rm B_{1g}$ $2\Delta$ peak
shifts with increasing doping to lower energies
from 570 cm$^{-1}$ in $\rm Y_{0.9}Pr_{0.1}$ and 
390 cm$^{-1}$ in Y-123 
down to 300 cm$^{-1}$ in $\rm Y_{0.95}Ca_{0.05}$.
Compared to fully oxygenated single crystals the energy of the  
$\rm B_{1g}$ $2\Delta$ peak in the film Y-123 is approximately 10 \% lower.\cite{Chen1993}
This indicates a higher doping level in the film compared to the 
single crystals. 
The increased doping level could result from additional oxygen which may enter 
the film at grain- or twin-boundaries.
These boundaries appear  
with much higher densities in the films than in the single crystals 
as a result of the lattice mismatch to the substrate. 

To give an example we show in 
Fig.~\ref{figFit}(a) the efficiency $I_{0}(\omega)$ of the 
film $\rm Y_{0.95}Ca_{0.05}$ measured at 18 K. 
Also shown is a fit to the spectrum where the $\rm B_{1g}$ phonon and 
the Ba mode are described according to Eq. (\ref{eqRamInt},$R_0$), 
and Lorentzians are taken for the O(4) mode and 
the weak feature around 600 cm$^{-1}$.
Subtracting all phonons from the spectrum we obtain the electronic 
background $\varrho_*(\omega)$ which is shown in Fig.~\ref{figFit}(b) 
together with the description according to Eq. (\ref{eqRho}) used in the fit.
The real part of the electronic background $R_*(\omega)$ is 
shown in Fig.~\ref{figFit}(c). 
In order to obtain $R_*(\omega)$ we have performed a numerical Hilbert 
transformation of $\varrho_*(\omega)$.
For the transformation the measured spectrum is taken as constant 
for high frequencies up to $\omega_{cut}$,
and is interpolated to zero intensity at $\omega=0$; for negative 
frequencies the antisymmetry of $\varrho_*(\omega)$ has been 
used.
Whereas a good description of the spectrum can also be obtained with 
a simpler model where the real part of the response function is 
considered constant,\cite{Chen1993}
our approach simultaneously yields a good description of $R_*(\omega)$
which is important when phonon self-energy effects are 
investigated.
The temperature dependencies of the $\rm B_{1g}$ $2\Delta$ peaks
$\omega_{2\Delta}$ of the films are 
shown in the inset of Fig. \ref{figFit} where a typical phonon 
frequency is indicated as a dashed horizontal line. 
Whereas the peak is always below 
the phonon frequency in the Ca-doped 
film it is always above the phonon in the Pr-doped film. 
In the film Y-123 the peak passes the phonon around 60 K.
In all films only weak temperature dependencies are observed in 
agreement with other Raman studies.\cite{Boekholt1991}
With increasing temperature the intensities of the 
background peaks decrease, becoming unresolvable above $T_{c}$.

The temperature dependencies of the fit parameters of the $\rm B_{1g}$ 
phonons are shown in Fig.~\ref{figTemp}.
Beside the bare phonon parameters and the self-energy contributions 
at $\omega=\omega_p$ also the renormalized frequencies 
$\omega_{\nu}(\omega_p)$ and linewidths 
$\gamma(\omega_p)$ 
are depicted which facilitates a comparison with previous data obtained 
with simpler Fano models.
Solid lines are fits to the anharmonic decay for both 
models,\cite{intensity} and anomalies below $T_c$ are 
discussed with respect to these fits.
Looking at the renormalized phonon parameters, we observe 
a behavior which is in agreement with the relative 
positions of $\rm B_{1g}$ $2\Delta$ peak and phonon energy:
We find a softening in $\rm Y_{0.9}Pr_{0.1}$ 
and Y-123 where $\omega_{2\Delta}(T\rightarrow 0)>\omega_p$ 
and a hardening in $\rm Y_{0.95}Ca_{0.05}$ where 
$\omega_{2\Delta}(T\rightarrow 0)<\omega_p$ (see Fig. \ref{figFit}). 
In those two films where the $\rm B_{1g}$ $2\Delta$ peaks evolve in the vicinity of 
the phonon we find strong broadenings of $\approx 5$ cm$^{-1}$  
whereas a sharpening of $\approx 2$ cm$^{-1}$ occurs in 
$\rm Y_{0.9}Pr_{0.1}$.
Our fit procedure now allows us to identify the self-energy effects 
which originate from the background.
For all films a convincing correlation exists 
between the changes of the linewidth appearing below $T_c$ and the varying 
self-energy contributions $\varrho_*(\omega_p)/C$.
It turns out that the sharpening in $\rm Y_{0.9}Pr_{0.1}$ is a 
consequence of a reduced effective coupling 
$g_{\mathrm{B_{1g}}}^2\propto 1/C$ at low 
temperatures, as the background intensity at the 
phonon position $\varrho_*(\omega_p)$ varies hardly with temperature 
(see Fig. \ref{figDopingRaman}).
The broadenings of the other films on the other hand are almost entirely 
described by increasing intensities $\varrho_*(\omega_p)$ accompanied 
by increasing effective couplings.
However, looking at the bare phonon linewidths $\Gamma$ some 
anomalies remain, i.e. a slight sharpening of $\approx 1.5$ cm$^{-1}$ 
and a slight broadening of $\approx 1$ cm$^{-1}$ in 
$\rm Y_{0.9}Pr_{0.1}$ and $\rm Y_{0.95}Ca_{0.05}$, respectively, as 
well as an intermediate broadening of $\approx 1.5$ cm$^{-1}$ in Y-123 
appearing around 60 K.
Regarding the frequency effects even stronger anomalies are observed. 
Whereas we find a good explanation of the softening in Y-123 as a 
result of the varying contribution $R_*(\omega_p)/C$, 
the bare phonon frequencies $\omega_p$ in the other films differ 
significantly from an anharmonic behavior, showing pronounced 
softening or hardening in $\rm Y_{0.9}Pr_{0.1}$ or 
$\rm Y_{0.95}Ca_{0.05}$, respectively.
The appearance of such an anomaly in the Pr-doped film is 
related to the fact that 
only a small redistribution is observed, the 
anomaly in the Ca-doped film, however, is unexpected.

Finally, we turn to the lowest row of 
Fig. \ref{figTemp} in which
beside the total intensities $I_{tot}$ also the bare phonon 
intensities $I_{phon}$ are given.
Even though increasing intensities are observed below $T_c$
in both films which exhibit 
$\rm B_{1g}$ $2\Delta$ peaks above the phonon energy, they have different origins.
Whereas the increase in $\rm Y_{0.9}Pr_{0.1}$ can be assigned to the 
increasing values of $R_0$ which compensate the decreasing 
effective coupling, they are a result of an increasing 
background contribution $R_*(\omega_p)$ accompanied by an increasing
effective coupling in the film Y-123.
No intensity anomaly is observed in the film $\rm Y_{0.95}Ca_{0.05}$ 
where independent of the temperature the $\rm B_{1g}$ $2\Delta$ 
peak is below the phonon energy.
A superconductivity-induced increase of the interband contribution 
to the Raman intensity of the $\rm B_{1g}$ phonon has been 
obtained in a theoretical treatment of the scattering 
process by Sherman {\em et al.}\cite{Sherman1995}
The latter authors showed that the increase will depend on the doping 
level, becoming stronger in underdoped samples.
This appears to be in 
agreement with our data.
Our results, however, also indicate that there is a particular correlation
between the intensity anomaly due to the interband 
contribution $R_0$ and the relative energies of 
phonon and $\rm B_{1g}$ $2\Delta$ peak which has not been observed before.
A further discussion of this correlation will be presented in 
a forthcoming publication.\cite{Bock1999rev}

\section{discussion and conclusions}
\label{sec:dis}

The remaining anomalies of the bare phonon parameters show that 
additional interaction effects are present which modify 
Eqs. (\ref{eqResponse}) and (\ref{eqRamInt},$R_0$).
In the following we will discuss two possible origins of the 
observed anomalies: 
(i) The assumption that the phonon interacts with the entire 
measured background is wrong; 
(ii) An additional excitation exists which 
renormalizes the phonon but is not Raman-active.

(i): If the electronic continuum consists of several contributions 
interacting only partly with the phonon, the applied fitting 
procedure would not allow us to obtain the proper bare phonon parameters. 
In fact, in the antiferromagnetic phase $\rm YBa_2Cu_3O_6$, in 
which intraband 
excitations have vanished, one still observes a background on which 
the $\rm B_{1g}$ phonon is superimposed.\cite{Burns1991} 
The symmetric appearance of the 
phonon indicates the absence of coupling to that background,
which can more or less unambiguously be related to 
two-magnon scattering.\cite{Lyons1988,Knoll1990}
At higher doping levels the two-magnon intensity 
decreases.\cite{Blumberg1994,Blumberg1997}
However, in previous works we have shown that in 
low up to slightly overdoped superconducting 
$\rm Y_{1-x}Pr_xBa_2Cu_3O_7$ (Ref. \onlinecite{Ruebhausen1997}) and 
$\rm Bi_2Sr_2CaCu_2O_{8+\delta}$ (Ref. \onlinecite{Ruebhausen1998}) 
compounds the magnetic scattering in $\rm B_{1g}$ symmetry 
becomes amplified below $T_c$ having significant spectral weight at 
low frequencies.
The observation that the effective coupling decreases below $T_c$ in 
$\rm Y_{0.9}Pr_{0.1}$ is thus in agreement with a strengthening of magnetic 
excitations in that temperature range. 
However, studying $\rm Bi_2Sr_2CaCu_2O_{8+\delta}$ single 
crystals\cite{Ruebhausen1998D} 
we have seen recently that the superconductivity-induced changes 
at high energy transfers vanish at very high doping levels. 
Therefore, the influence of the magnetic scattering should decrease 
with increasing doping.
The similar strength of the anomalies in the overdoped film 
$\rm Y_{0.95}Ca_{0.05}$ can thus not be related to the existence of 
magnetic excitations which presumably are strongly overdamped at 
this doping level.
Therefore, our observation might have a different origin at higher doping 
levels.

(ii): In the second scenario anomalies in the underdoped as well as in 
the overdoped regime can be explained by assuming the presence of 
an additional not Raman-active excitation. 
The energy of this excitation must be strongly related to the gap 
value in order to explain the intermediate broadening in Y-123 
which appears in the temperature range where the measured 
$\rm B_{1g}$ $2\Delta$ peak passes the phonon frequency.
In fact, the doping dependence of such a broadening in several 
$\rm YBa_2Cu_3O_{7-\delta}$ single crystals with different oxygen 
contents\cite{Altendorf1993}
agrees well with the doping dependence of the $\rm B_{1g}$ $2\Delta$ 
peak obtained on the same crystals.\cite{Chen1993} 
The intermediate broadening cannot be attributed to the 
Raman-active gap excitations
as $\varrho_*(\omega_p)/C$ shows a monotonic behavior at the 
considered temperatures within the error bars.
The small temperature interval of the broadening indicates that the 
imaginary part of the response function associated with this 
excitation has a small extension in frequency space of roughly 
$\pm 20$ cm$^{-1}$ as $\omega_{2\Delta}$ varies only slightly below $T_c$. 
$2\Delta$ peaks of this sharpness have indeed not been observed in 
Raman spectra of cuprates so far.
Even though the imaginary part of the response function has a rather 
small width, the real part will extend to a somewhat 
broader frequency range and could therefore be responsible for the 
anomalous softenings and hardenings in
$\rm Y_{0.9}Pr_{0.1}$ and $\rm Y_{0.95}Ca_{0.05}$.
In this picture the almost anharmonic behavior of the bare phonon 
frequency in Y-123 is not surprising as the real part changes sign 
when the gap passes the phonon.

The origin of the additional excitation which 
renormalizes the phonon but does not show up in the Raman spectra 
could possibly be found in a not Raman-active gap excitation.
As seen before\cite{Nemetschek1997} and shown in this work 
the gap features rapidly loose their strength at 
lower doping levels.
This cannot be understood on the basis of the standard
theory\cite{Klein1984} where a metal-like band picture is used to 
describe the response in the superconducting state.
It has therefore been suggested that vertex corrections due to the strong 
correlations should be considered in order to understand the 
additional screening which suppresses the gap features in the Raman 
process.\cite{Manske1998,Ruebhausen1998D}
This work provides some evidence that screening 
of the $\rm B_{1g}$ Raman response may not only be
present around or below the optimum doping level where the cuprates 
may be regarded as doped antiferromagnets, but also above this level 
when the cuprates start to evolve into a metallic state.

In conclusion, we present a model that allows us to study background 
redistributions and related phonon self-energy effects in 
$\rm Y_{1-x}(Pr,Ca)_xBa_2Cu_3O_7$ films simultaneously.
In this model we use a phenomenological description of the 
electronic background and find that this approach enables us to 
explain various self-energy effects.
However, some anomalies remain.
We discuss whether these anomalies result from the fact that the 
observed background consists in a subtle way of electronic as well as 
magnetic excitations, or whether they also indicate
that another not Raman-active 
excitation, which we suggest to originate from the gap itself, 
renormalizes the phonon.

\acknowledgements

The authors thank M. K\"all, M.V. Klein, D. Manske and U. 
Merkt for stimulating discussions.
S.O., C.T.R., and M.R. acknowledge grants of the German Science 
Foundation via the Graduiertenkolleg 
``Physik nanostrukturierter Festk\"orper''.

\begin{figure}
\caption{Feynman diagram for phononic Raman scattering, where
  $\gamma(\mathbf{k})$ and $g(\mathbf{k})$ are the electron-photon and 
  electron-phonon vertices, and
  solid and wavy lines the electron and phonon propagator.
  The outer dashed lines correspond to the photons.}
\label{figFeynman}
\end{figure}
\begin{figure}
\caption{Raman efficiencies in $\rm B_{1g}$ symmetry  
  of the three investigated films at
  $T=18\,\rm K$ (solid lines) and $152\,\rm K$ (dots).  
  The insets show the electronic backgrounds after subtraction of 
  the phonons.
  The efficiencies are given in the same units.}
\label{figDopingRaman}
\end{figure}
\begin{figure}
\caption{(a) Raman efficiency $I_0(\omega)$ in $\rm B_{1g}$ symmetry 
  of the film $\rm Y_{0.95}Ca_{0.05}$ at $T=18\,\rm K$ (open 
  circles),
  and fit result (solid line). 
  (b) Imaginary part $\varrho_*(\omega)$ and (c) numerically determined 
  real part $R_*(\omega)$ of the electronic background (open circles) 
  obtained after subtraction of the phonons, 
  and numerical descriptions used in the fit (solid lines). 
  Efficiencies and real part are given in the same units.
  Arrows indicate the energy 
  of the $2\Delta$ peak given by the fit parameter $\omega_{2\Delta}$.
  The inset shows the 
  energies $\omega_{2\Delta}$ versus temperature for the three 
  investigated films as indicated. The horizontal dashed line represents 
  a typical energy of the $\rm B_{1g}$ phonon of 345 cm$^{-1}$.
  }
\label{figFit}
\end{figure}
\begin{figure}
\caption{Temperature dependence of the fit parameters for the $\rm B_{1g}$ 
  phonon. Closed circles represent the bare phonon parameters  
  $\omega_p$, $\Gamma$, and $I_{phon}$, 
  crosses the self-energy contributions 
  $R_*(\omega_p)/C $ and $\varrho_*(\omega_p)/ C$, and 
  open circles the calculated renormalized values 
  $\omega_{\nu }(\omega_p)$, $\gamma(\omega_p)$, and $I_{tot}$.
  Solid lines are fits according to an anharmonic decay.
  The vertical dotted lines indicate the respective transition 
  temperatures of the films.
  Marker sizes represent the vertical accuracies.}
\label{figTemp}
\end{figure}


\begin{references}
\bibitem{KKRvonTanH} The Hilbert transformed incoherent 
   contribution is: \\ 
   $R_{inc}(\omega)= 
   I_{\infty}\left[0.5 + \frac{2}{\pi}\, {\rm ln}(\omega_{cut}/\omega_T) +
                           \sqrt{2} \, {\rm artanh}(p-1)\right]$, with
  $p=\tanh(\omega/\omega_T)/(\omega/\omega_T)$.
\end{references}
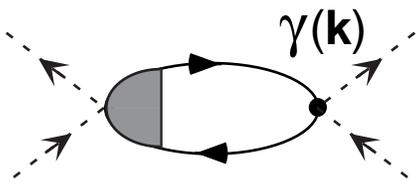
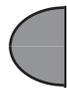
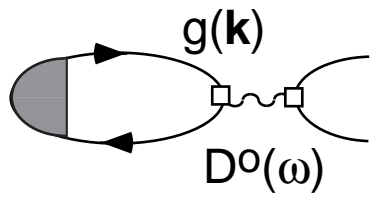
\end{document}